\documentclass{article}

\usepackage{PRIMEarxiv}

\usepackage[utf8]{inputenc} 
\usepackage[T1]{fontenc}    
\usepackage{hyperref}       
\usepackage{url}            
\usepackage{booktabs}       
\usepackage{amsfonts}       
\usepackage{nicefrac}       
\usepackage{microtype}      
\usepackage{lipsum}
\usepackage{fancyhdr}       
\usepackage{graphicx}       
\graphicspath{{media/}}     
\usepackage{natbib}
\usepackage{amsmath} 
\usepackage{multirow} 

\usepackage{natbib}

\title{Multi-category solar radio burst detection based on task-aligned one-stage object detection model
\thanks{\textit{\underline{Citation}}: 
\textbf{Mingming Wang et al. Multi-category solar radio burst detection based on task-aligned one-stage object detection model. Accepted for publication in Astrophysics and Space Science.}} 
}

\author{
  Mingming Wang \\
  School of Information Science and Engineering \\
  Yunnan University\\
  Kunming, China\\
   \And
  Guowu Yuan* \\
 School of Information Science and Engineering \\
  Yunnan University\\
  Kunming, China\\
  \texttt{gwyuan@ynu.edu.cn} \\
   \And
  Hailan He \\
 School of Information Science and Engineering \\
  Yunnan University\\
  Kunming, China\\
   \And
  Chengming Tan \\
 State Key Laboratory of Space Weather \\
  National Space Science Center\\
  Beijing, China\\
   \And
 Hao Wu  \\
 School of Information Science and Engineering \\
  Yunnan University\\
  Kunming, China\\
   \And
 Hao Zhou \\
 School of Information Science and Engineering \\
  Yunnan University\\
  Kunming, China\\
  }

\begin{document}


\maketitle

\begin{abstract}
Accurate identification of solar radio bursts (SRBs) is essential for advancing research in solar physics and predicting space weather. However, the majority of current studies mainly concentrate on detecting whether SRBs are present or absent, often focusing on only one particular type of burst. Moreover, the neural network models used for SRB detection are typically complex, involving a large number of parameters, which results in slower processing speeds. This study establishes a dataset encompassing Type II, Type III, Type IIIs, Type IV, and Type V SRBs collected from e-CALLISTO, including 8,752 SRB spectrum images and achieving annotations for 10,822 SRBs. We propose a multi-category SRB detection model based on task-aligned one-stage object detection (TOOD). TOOD can solve the problem of inconsistent predictions in classification and localization tasks, and it improves the detection recall rate. This model aligns classification and localization tasks and optimizes the neck network by incorporating a channel attention mechanism. This model achieves higher recall and accuracy with fewer parameters. This model can accurately detect five types of SBRs. The experimental results show that the model achieved an accuracy of 79.9\% (AP50) and a recall rate of 95.1\% on the SBRs dataset. A higher recall rate than other models means fewer SRBs are missed in automatic detection. The model we propose has the potential to make a substantial impact on solar physics research and space weather studies. Additionally, the findings in this paper could provide valuable insights for processing other small-sample astronomical datasets.The source code and data is available at \href{https://github.com/onewangqianqian/MobileNetVitv2-TOOD.git}{https://github.com/onewangqianqian/MobileNetVitv2-TOOD.git.}
\end{abstract}



\section{Introduction}
Solar radio bursts (SRBs) directly reflect the activity of high-energy particles from the Sun. Monitoring SRBs helps diagnose the physical state of solar active regions \cite{ref1}. This includes the background plasma, non-thermal electrons, and the evolution and interactions of the magnetic fields. Solar activity occasionally manifests as intense explosive events such as coronal mass ejections and solar flares. These phenomena eject substantial clouds of high-temperature plasma, release enormous amounts of energy, and accelerate a significant number of high-energy charged particles \cite{ref2}. The radiation of high-energy particles from the Sun poses serious threats to various high-tech systems, including spacecraft, space stations, crewed space missions, satellite communications, satellite navigation, and large electrical grids on the ground, thereby constituting hazardous space weather events \cite{ref3}.

According to the distinct characteristics of SRBs detected by the spectrometer across various frequency ranges, meter-wave SRBs are typically categorized into five types: Type I (which includes I-type and enhanced radiation), Type II, Type III, Type IV, and Type V \cite{ref2,ref4,ref5}. Each type of SRB event is associated with specific solar activity phenomena and is also related to various phenomena in the near-earth space environment \cite{ref2}. Type I SRBs are characterized by gradual, broadband, continuous solar storms that can last from several hours to several days. Type II SRBs display narrowband radiation with a frequency that drifts slowly over time, suggesting the presence of coronal shock waves. In contrast, Type III SRBs are distinguished by a rapid frequency drift, with frequencies decreasing sharply over time. These bursts indicate the movement of high-energy electron beams through the corona and are crucial for understanding energy release and particle acceleration during solar flares or coronal mass ejections. Type IV SRBs consist of continuous, wideband radiation generated by synchrotron radiation from high-energy electrons spiraling within magnetic fields. Finally, Type V SRBs have a broad frequency range and are relatively rare. Typically observed after Type III SRBs, they reflect strong scattering of high-energy electrons by coronal waves \cite{ref6}.

The real-time detection and classification of SRBs are essential in solar physics research and space weather forecasting. Timely and accurate space weather alerts are crucial for ensuring the safety of aerospace operations, satellite communications, and the functioning of electrical grids. Traditionally, detection and classification are done manually \cite{ref7}, requiring operators to have specific astronomical knowledge and using a lot of human resources. This method cannot process data in real-time and may miss events. Thus, developing efficient methods for detecting and classifying SRBs is critical for advancing solar physics research, and space weather forecasting \cite{ref8}.
\section{Related Work}

Deep learning has become a crucial tool in the processing of astronomical data, aiding researchers in handling large datasets, extracting meaningful information, and enhancing data processing efficiency, all while minimizing manual effort \cite{ref9}. As solar radio spectrometers continue to advance, the demand for efficient, real-time processing of extensive observational data to derive valuable insights has become increasingly urgent. Researchers have recently utilized deep learning techniques to classify and detect solar radio spectrogram images.

One of the research trends in solar radio spectrogram images is the classification and detection of the fine structures of SRBs. This involves utilizing advanced deep-learning models such as the RCNN series, SSD, and YOLO series. Researchers seek to enhance these models by leveraging the unique characteristics of solar radio spectrogram to develop more reliable models specifically optimized for this data type.

Most contemporary research is centered on detecting a specific type of SRB, particularly Type III SRBs, or on classifying the presence of these bursts. In the domain of solar radio spectrogram image classification, reference \cite{ref10} employs VGG16 and transfer learning to classify solar radio spectrogram images, avoiding overfitting due to the small sample size of solar radio spectrogram datasets. Reference \cite{ref11} uses YOLOv2 for detection and classification of Type III SRBs on a dataset solely composed of simulated Type III bursts. In solar radio spectrogram images, the horizontal axis represents time, while the vertical axis represents frequency, capturing both temporal and spatial characteristics. To leverage this feature, reference \cite{ref12} proposes a classification method for solar radio spectrograms based on Long Short-Term Memory (LSTM) neural networks. Reference \cite{ref13} introduces a Convolutional LSTM (ConvLSTM) model that incorporates a spatial attention mechanism to learn image features tailored to the characteristics of solar radio spectrograms. Reference \cite{ref14} introduces a hybrid architecture that integrates convolutional layers and memory units, enabling the simultaneous extraction of both frequency structure and time series features. This approach improves the model's sensitivity to subtle features within spectrograms, aiding in more accurate classification. Reference \cite{ref15} proposes a self-supervised learning method for classifying solar radio spectrograms, where the model is pre-trained on other datasets using a self-masking approach and fine-tuned on solar radio spectrogram datasets, achieving accuracy comparable to supervised learning. Lastly, reference \cite{ref16} introduces a classification method for solar radio spectrograms based on Swin Transformer, aiming to achieve classification with a lower number of model parameters.

Additionally, researchers have also studied the classification and detection of multiple types of SRBs. Reference \cite{ref17} introduces a small-sample object detection method utilizing transfer learning to classify Type II, Type III, and Type IV SRB events. Due to the low-probability nature of SRBs, the distribution of spectrogram data samples is highly imbalanced, with significantly fewer burst spectrogram images compared to quiet Sun spectrogram images. To address this imbalance, some researchers have employed Generative Adversarial Networks (GANs) to generate synthetic sample images. Reference \cite{ref18} proposes a conditional information-based Deep Convolutional Generative Adversarial Network (C-DCGAN) using solar radio spectroscopic data from the Culgoora and Learmonth observatories. This approach aims to automatically classify five types of SRB events, helping to mitigate the overfitting issues associated with limited data samples. Reference \cite{ref19} employs the one-stage object detector SSDLite for the rapid detection of SRBs events. However, this method uses two independent branches to perform target classification and localization in parallel. This may result in ineffective interaction between the two tasks, potentially leading to inconsistencies between the best classification and localization anchors. Reference \cite{ref31} proposes a multicategory SRB detection method based on YOLOv8. However, the detection accuracy remains relatively low.

However, research on simultaneous automatic recognition and detection of multiple types of SRBs is still needed. There is also a lack of publicly available SRB spectrogram datasets that include various types of SRBs.

\section{Main contributions}

Deep learning can improve the accuracy of detecting and identifying SRBs. However, current research still faces several issues: (1) Most studies focus on detecting whether a burst occurs or identifying Type III SRBs, lacking comprehensive classification and localization research for other SRBs. (2) The recall rate of the previous model was insufficient, resulting in many missed SRBs. Automatic object detection models are usually used as the initial filtering for SRBs, followed by manual confirmation to determine the results. So, automatic detection aims to detect possible SRBs as much as possible, which requires the highest recall rate. (3) Convolutional neural networks have limited receptive fields, hindering their ability to combine information from the time- and frequency-domain of solar radio spectra effectively.

Designing an efficient and accurate detection model for multi-type SRBs is necessary. This research aims to improve a deep learning-based object detection model to detect SRB events automatically. We will compare and analyze various models and the current research on detecting SRBs, leading to the study's results. The main contributions of this paper are as follows:

(1) This paper establishes a dataset collected from e-CALLISTO, including Type II, Type III, Type III-s, Type IV, and Type V SRB spectral images. We have spent much time annotating these SRB spectral images. This database will be shared with other researchers for related research.

(2) This paper proposes an SRBs detection model based on task-aligned object detection (TOOD). The proposed model enhances the detection recall rate by aligning the classification and localization tasks and incorporating channel attention to optimize the neck network. Compared to the previous model, this improved model minimizes the number of missed SBRs while ensuring high accuracy.

\section{Dataset construction and augmentation}

\subsection{Dataset construction}
With the development of solar physics research, many high-performance observational devices have been established worldwide, such as the solar radio spectrometer built at the Nancy Observatory in France,  the MingantU SpEctral Radioheliograph (MUSER) developed by the National Space Science Center in China, and the e-CALLISTO (Compound Astronomical Low frequency Low cost Instrument for Spectroscopy and Transportable Observatory) \cite{ref20,ref21,ref22}. These spectrometers play a significant role in monitoring solar bursts.
   
The e-CALLISTO network consists of CALLISTO spectrometers distributed globally, capable of continuously observing solar radio spectra 24 hours a day throughout the year, monitoring emissions in the meter and decimeter wave bands \cite{ref25}. The publicly available data from the e-CALLISTO network is comprehensive, which allows this study to utilize the publicly accessible solar radio spectrogram data from e-CALLISTO to establish a dataset of SRBs.

The data obtained from e-CALLISTO is in Flexible Image Transport System (FITS) file format. These CALLISTO instruments automatically collect and transmit data daily via the internet. All site data is gathered in a centralized database with a public web interface for browsing and retrieving data (www.e-callisto.org).

This study uses the publicly available SRB spectrum FITS data from the e-CALLISTO system, which is parsed into solar radio spectrogram images using the pyCALLISTO software. We have established a dataset containing five types of SRB events (Type II, Type III, Type III-s, Type IV, and Type V). As e-CALLISTO simultaneously provides observational data from different stations worldwide, the dataset is constructed with spectrogram images with relatively low noise, requiring minimal denoising. Thus, only background subtraction was performed before labeling the spectrogram images.

We use the information provided in the records of SRB events in e-CALLISTO, and the collected spectrogram images are annotated for location and category information using the Labelme tool. This creates an object detection dataset formatted similarly to the COCO dataset. The dataset is split into training and validation sets in a 7:3 ratio, facilitating the training and evaluation of subsequent object detection learning models. Table \ref{table1} presents the instances for each type in the training and validation sets. In Table \ref{table1},  the number of SRB instances exceeds the number of spectrogram images due to multiple SRB events occurring within a single spectrogram image.

   \begin{table*}[h!]
      \caption{Statistics of the dataset's sample for each solar radio burst (SRB).}
         \label{table1}
         \centering
         \resizebox{0.9\linewidth}{!}{
         \begin{tabular}{cccccccc}
            \hline
            \hline
              &  Number of  Images & Number of  Instances & II & III & IIIs & IV & V \\
            \hline
            Train set & 6126 & 7584 & 338 & 5736 & 1296 & 47 & 167    \\
            Val set & 2626 & 3238 & 126 & 2474 & 561 & 18 & 70 \\
            \hline
         \end{tabular}
        }
   \end{table*}

The CALLISTO network provides spectrogram images with a standardized duration of 15 minutes. Type I solar radio bursts are characterized by sequences of brief, intermittent bursts superimposed on a stationary or gradually varying continuum background. These eruptive processes typically span multiple consecutive spectrogram images, making accurate annotation challenging when based on individual 15-minute spectral observations. Consequently, the current study excluded Type I bursts from the experimental dataset. However, we anticipate that future investigations employing spectrogram concatenation techniques will enable effective annotation of Type I burst phenomena. Moreover, the occurrence frequency of Type IV SRBs is lower than that of other types, resulting in a much smaller number of Type IV SRBs in this dataset. Based on the classification and burst catalog available on the e-CALLISTO website, Type III SRBs are further categorized into two subtypes: individual Type III SRBs and grouped Type III SRBs (referred to as Type III-s).

\subsection{Data augmentation}
\begin{figure*}[h!]
    \centering
    \includegraphics[width=1.0\hsize]{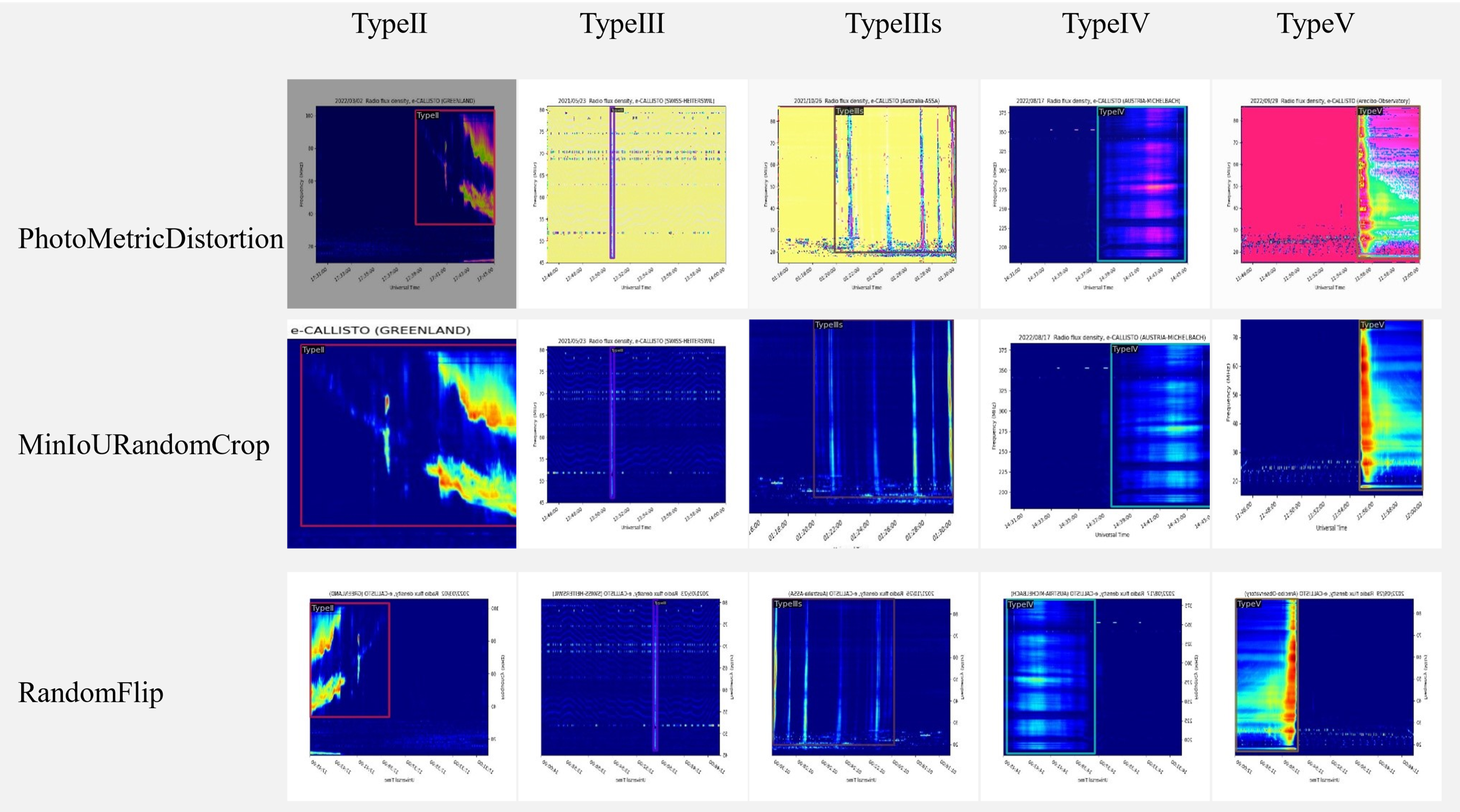}
    \caption{Data Augmentation for Solar Radio Spectrogram Images.}
    \label{fig:fig1}
\end{figure*}

Given the limited size of the dataset and the uniform background across different classes, this study utilizes commonly employed data augmentation techniques to expand the training samples and improve the robustness and accuracy of the model. The augmentation methods applied include random flipping (RandomFlip), minimum intersection over union random cropping (MinIoURandomCrop), and photometric distortion (PhotoMetricDistortion, PMD), as shown in Figure \ref{fig:fig1}.

By integrating various data augmentation methods in Figure \ref{fig:fig1}, a more diverse and enriched training dataset of SRB spectrograms can be created. A greater variety of training samples can strengthen the model's generalization ability and robustness, enhancing performance and effectiveness. The augmented images are exclusively used for training, while only real spectrogram images are employed for testing.

\section{Methods}

Artificial intelligence can improve efficiency, but it will take many years to replace human work completely. In this application, automatic detection is usually used as the initial filtering for SRBs, followed by manual confirmation to determine the results. This can significantly reduce the cost of manual detection. So, automatic detection aims to detect possible SRBs as much as possible, which requires the highest recall rate.

This study introduces task-aligned one-stage object detection (TOOD) to improve the detection recall rate. TOOD can solve the problem of inconsistent predictions in classification and localization tasks and improve the detection recall rate. In addition, we optimize the neck network by incorporating a channel attention mechanism in TOOD. The improved model aligns classification and localization tasks and can achieve higher recall and accuracy with fewer parameters.

\subsection{Proposed detection model}
Object detection is typically considered a multi-task learning problem that jointly optimizes target classification and localization. However, the classification task aims to learn discriminative features that focus on the key or salient parts of the target. In contrast, the localization task is dedicated to accurately determining the boundaries of the target. The different learning mechanisms of classification and localization tasks may lead to discrepancies in the feature space distribution learned by the two tasks \cite{ref23,ref24}. TOOD has a task-aligned head structure and an alignment-focused learning approach to more accurately align the classification and detection tasks.

\begin{figure*}[h!]
    \centering
    \includegraphics[width=1\hsize]{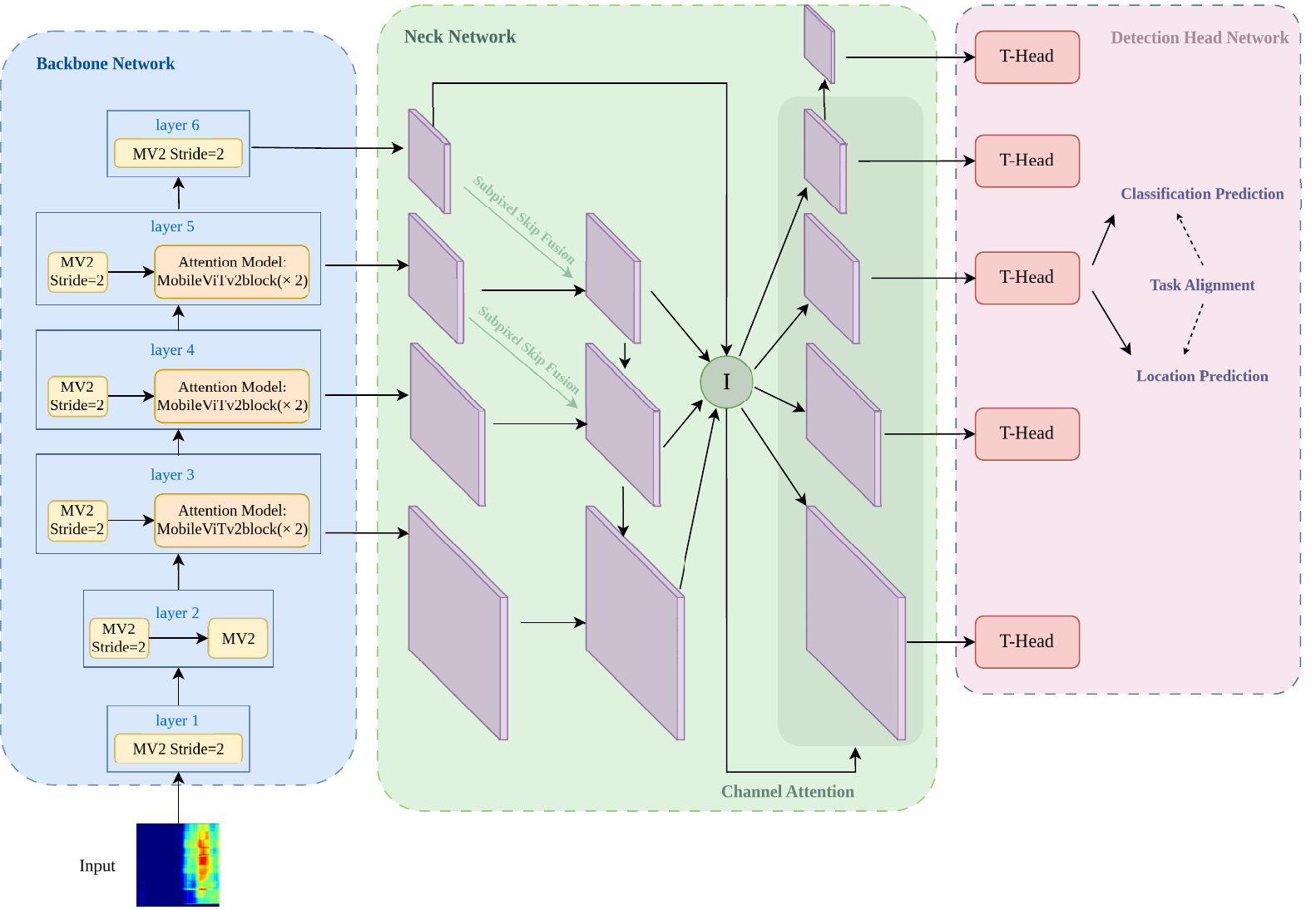}
    \caption{Overall structure of our proposed model}
    \label{fig:fig2}
\end{figure*}

TOOD \cite{ref23} is a one-stage object detector characterized by a structure consisting of a backbone network (ResNet), a neck network (FPN), and a task-aligned detection head. This paper proposes a model specifically for the practical needs of SRB detection. The overall structural diagram of the model is shown in Figure \ref{fig:fig2}. This model replaces the backbone network ResNet with MobileViTv2 and employs a channel-attention-enhanced FPNLight to learn and merge spectral features. Furthermore, it incorporates the task alignment concept from TOOD to enhance the non-aligned independent detection branches.

In Figure \ref{fig:fig2}, MobileViTv2 replaces the backbone network ResNet. SRB spectrogram images are processed through the MobileViTv2 backbone to extract global temporal features and frequency domain features. Next, a channel-attention-enhanced FPNLight is employed to learn and merge the spectral features. FPNLight can integrate high-level semantic information with low-level spatial information, generating a consolidated feature map $ I $, which feeds the attention-guided features from each layer into the detection head network. Finally, the task alignment concept from TOOD is applied to improve the non-aligned independent detection branches. The task detection head, called T-Head, outputs classification and localization results. At the same time, the task-aligned learning objective function trains the model, aligning the optimal classification anchor points with the optimal localization anchor points.

Afterward, Section \ref{Sec5.2} introduces task-aligned one-stage object detection (TOOD), including task-aligned head (T-Head) and task-aligned learning (TAL). Section \ref{Sec5.3} introduces the improved parts of the backbone and neck network in Figure \ref{fig:fig2}, which are used for feature extraction and fusion.

\subsection{Task-aligned one-stage object detection (TOOD)} \label{Sec5.2}
One-stage object detectors provide more accurate classification and localization predictions through anchor points or anchor boxes located at the center of the target. However, this method has certain limitations: First, current one-stage object detectors parallelize target classification and localization through two independent branches, leading to a lack of effective interaction between tasks, which results in inconsistencies in predictions when performing tasks. Second, the widely used sample allocation schemes are task-agnostic, meaning that the spatial locations of the best localization anchors may not align well with the optimal classification anchors. This can lead to precise bounding boxes being suppressed \cite{ref23} by less accurate ones during the non-maximum suppression (NMS) process. To address these issues, TOOD proposes a novel task-aligned head (T-Head) structure, which enhances the interaction between the two tasks while preserving the features for classification and localization, further aligning the predictions. Additionally, TOOD introduces a task-aligned learning (TAL) method, which explicitly defines the distance between the optimal anchors of the two tasks by designing a task-aligned sample allocation scheme and related loss functions. The collaboration of T-Head and TAL enhances the consistency of classification and localization tasks.

\subsubsection{Task-aligned head (T-Head)}
\begin{figure*}[h!]
    \centering
    \includegraphics[width=1.0\hsize]{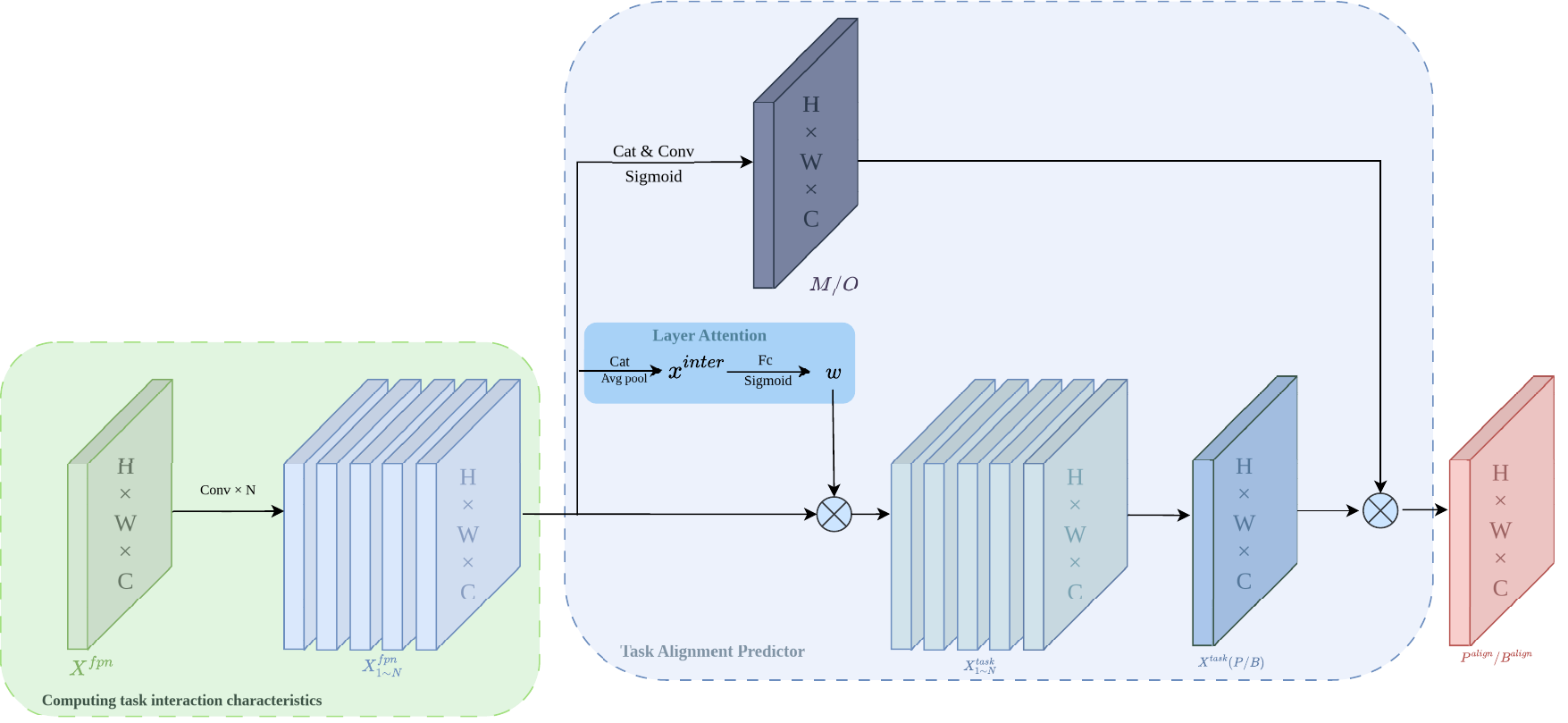}
    \caption{Structure of the task-aligned head (T-Head)}
    \label{fig:fig3}
\end{figure*}

Figure \ref{fig:fig3} shows that the T-Head consists of a feature extractor and a task-aligned predictor (TAP). The T-Head first computes task interaction features using $N$ consecutive convolutional layers to facilitate task interaction. It then utilizes the TAP to calculate task-specific features.

The Task-Aligned Predictor (TAP) addresses the feature conflict between classification and localization tasks in object detection. Through dynamic feature selection, spatial attention mechanisms, and prediction refinement, TAP explicitly aligns classification scores with localization accuracy to enhance task consistency. The mathematical formulation is expressed as follows:
\begin{eqnarray}
    \label{equ1}
    P^{\text{align}}(i,j,c) = \sqrt{P \cdot M} \\
    B^{\text{align}}(i,j,c) = B + \Delta O
\end{eqnarray}

\begin{eqnarray}
    \label{equ2}
    M = \sigma\left(\text{Conv}_2\left(\sigma\left(\text{Conv}_1(x^{\text{inter}})\right)\right)\right) \\
    O = \text{Conv}_4\left(\sigma\left(\text{Conv}_3(x^{\text{inter}})\right)\right)
\end{eqnarray}

Specifically, \(P^{\text{align}}\) represents the aligned classification features, obtained by combining the original classification features \(P\) with the spatial probability map \(M \in \mathbb{R}^{H \times W \times 1}\) through a specific operation. \(M\) is derived by applying two convolutional operations to the interaction features \(x^{\text{inter}}\) generated by the feature extractor. Here, \(\text{Conv}_1\) and \(\text{Conv}_2\) are 1×1 convolutions, followed by the \(\sigma\) (Sigmoid) function. \(B^{\text{align}}\) denotes the aligned localization features, computed from the original localization features \(B\) and the spatial offset map \(O \in \mathbb{R}^{H \times W \times 8}\). \(O\) is also obtained by applying two convolutional operations to the interaction features \(x^{\text{inter}}\), where \(\text{Conv}_3\) and \(\text{Conv}_4\) are 3×3 convolutions with an output channel size of 8.

\subsubsection{Task-aligned learning (TAL)}

The core idea of task-aligned learning (TAL) includes a sample assignment strategy based on task alignment metrics and a task alignment loss. The task alignment metric \( t \) measures the degree of alignment between detection and classification tasks, calculated as follows:

\begin{eqnarray}
    t = s^{\alpha} \times u^{\beta}
\end{eqnarray}
where, \( s \) and \( u \) represent the classification score and the IoU value, respectively, and \( \alpha \) and \( \beta \) control the influence of these two tasks on the anchor alignment metric. The task alignment metric \( t \) is integrated into the sample assignment strategy and the target loss function, dynamically improving the predictions at each anchor box or point.

The positive and negative sample assignment strategy for task alignment selects the top-k anchors with the largest \( t \) values as positive samples. In contrast, the remaining anchors are treated as negative samples. In the task alignment loss, the classification loss uses \( t \) as the positive sample label, while the localization loss weights the CIoU loss by \( t \). The total training loss for TAL is the sum of \( L_{cls} \) and \( L_{reg} \), calculated as follows:

\begin{eqnarray}
    L_{cls} & = & \sum_{i=1}^{N_{pos}} \left| \hat{t}_{j} - s_{j} \right|^{r} BCE\left( s_{j}, \hat{t}_{j} \right) + \sum_{i=1}^{N_{neg}} s_{j}^{r} BCE\left( s_{j}, 0 \right) \\
    L_{reg} & = & \sum_{i=1}^{N_{pos}} \hat{t}_{i} L_{CIoU}\left( b_{i}, \hat{b}_{i} \right)
\end{eqnarray}
where, \( \hat{t} \) is the normalized \( t \), where the maximum value of \( \hat{t} \) corresponds to the maximum IoU value for each instance, explicitly enhancing the classification score of aligned anchors while reducing the score for misaligned anchors (i.e., those with smaller \( t \) values). The classification loss is based on Focal Loss to address the imbalance between negative and positive samples during training, where \( \gamma \) is the focusing parameter. \( s \) represents the classification score, \( i \) refers to the \( i \)-th positive sample, and \( j \) corresponds to the \( j \)-th negative sample. The localization loss calculates the bounding box regression loss for each anchor with re-weighting, where \( b_{i} \) denotes the predicted bounding box and \( \hat{b}_{i} \) represents the ground truth bounding box.

\subsection{Feature extraction network} \label{Sec5.3}

TOOD \cite{ref23} employs a backbone residual network (ResNet) and a neck feature pyramid network (FPN) structure as feature extractors. However, a mobile vision Transformer (MobileViT) demonstrates better learning capabilities for spectral features, leading to using the computationally more efficient MobileViTv2 instead of the ResNet backbone. Additionally, the FPN suffers from information attenuation during the feature fusion process and an aliasing effect during cross-scale fusion. To address this, sub-pixel convolution is introduced to reduce channel information loss, and a channel attention mechanism is utilized to optimize the fused aliased features, thereby enhancing the discriminative capability of the features.

\subsubsection{MobileViTv2}
\begin{figure*}[h!]
    \centering
    \includegraphics[width=1.0\hsize]{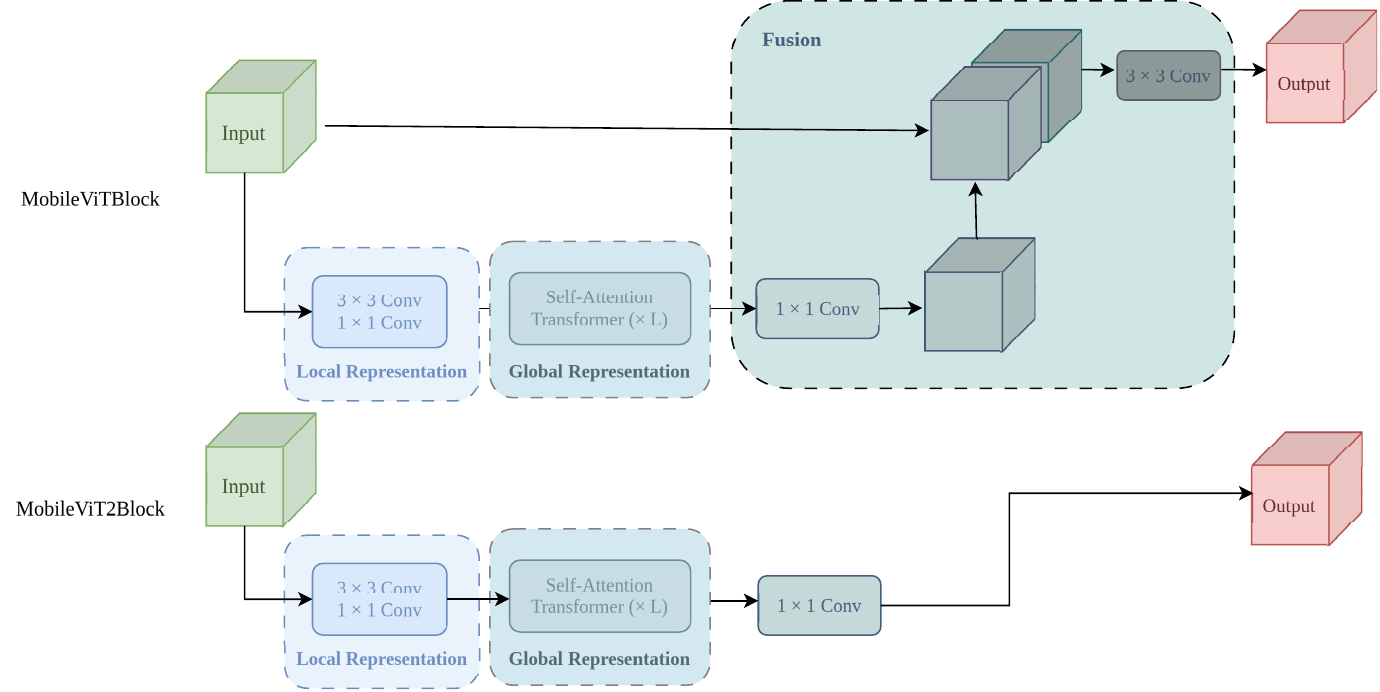}
    \caption{MobileViTv2Block and MobileViTBlock structure}
    \label{fig:fig4}
\end{figure*}

MobileViTv2 \cite{ref25} (Mobile vision Transformer with separable self-attention) is an improved version of MobileViT, designed to reduce the computational overhead and performance bottlenecks associated with the multi-head attention (MHA) operation in MobileViT. Figure \ref{fig:fig4} presents a structural comparison between them. MobileViTv2 eliminates the residual structure and the feature fusion component, transforming the convolutional neural network (CNN) into depth-separable convolution (DSC) operations within the local representation learning module to decrease computational cost. In the global representation structure, MobileViTv2Block employs separable attention instead of the multi-head attention used in MobileViT.

\begin{figure*}[h!]
    \centering
    \includegraphics[width=1.0\linewidth]{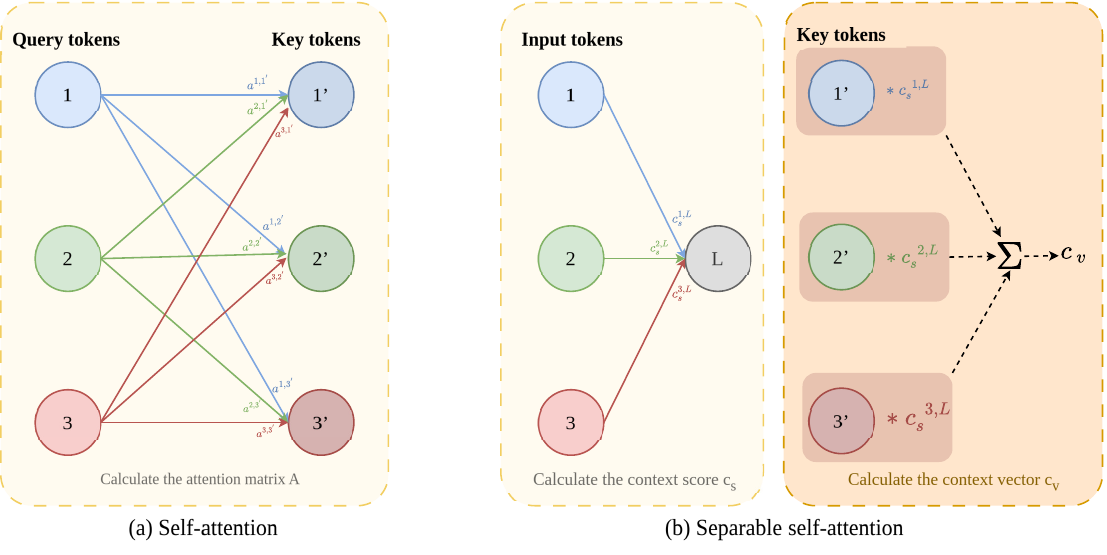}
    \caption{Self-attention and separable self-attention}
    \label{fig:fig5}
\end{figure*}

In MobileViT, the global representation module employs multi-head attention (MHA), which has a time complexity of \( O(K^2) \) for each vector. The computation of self-attention uses batch matrix multiplication, resulting in high network latency \cite{ref25}. In response, MobileViTv2 introduces separable self-attention with a time complexity of \( O(K) \), which encodes global information through two separate linear computations. As shown in Figure \ref{fig:fig5}(b), the core idea of separable attention is to compute context scores regarding the latent tokens \( L \) to re-weight the input tokens, producing a context vector that encodes global information. This method is implemented using element-wise operations (addition and multiplication), replacing MHA with two separate linear operations, thereby reducing self-attention complexity in the Transformer by a factor of \( K \).

\subsubsection{Feature pyramid network (FPN) enhanced by channel attention}

\begin{figure*}[h!]
    \centering
    \includegraphics[width=1.0\hsize]{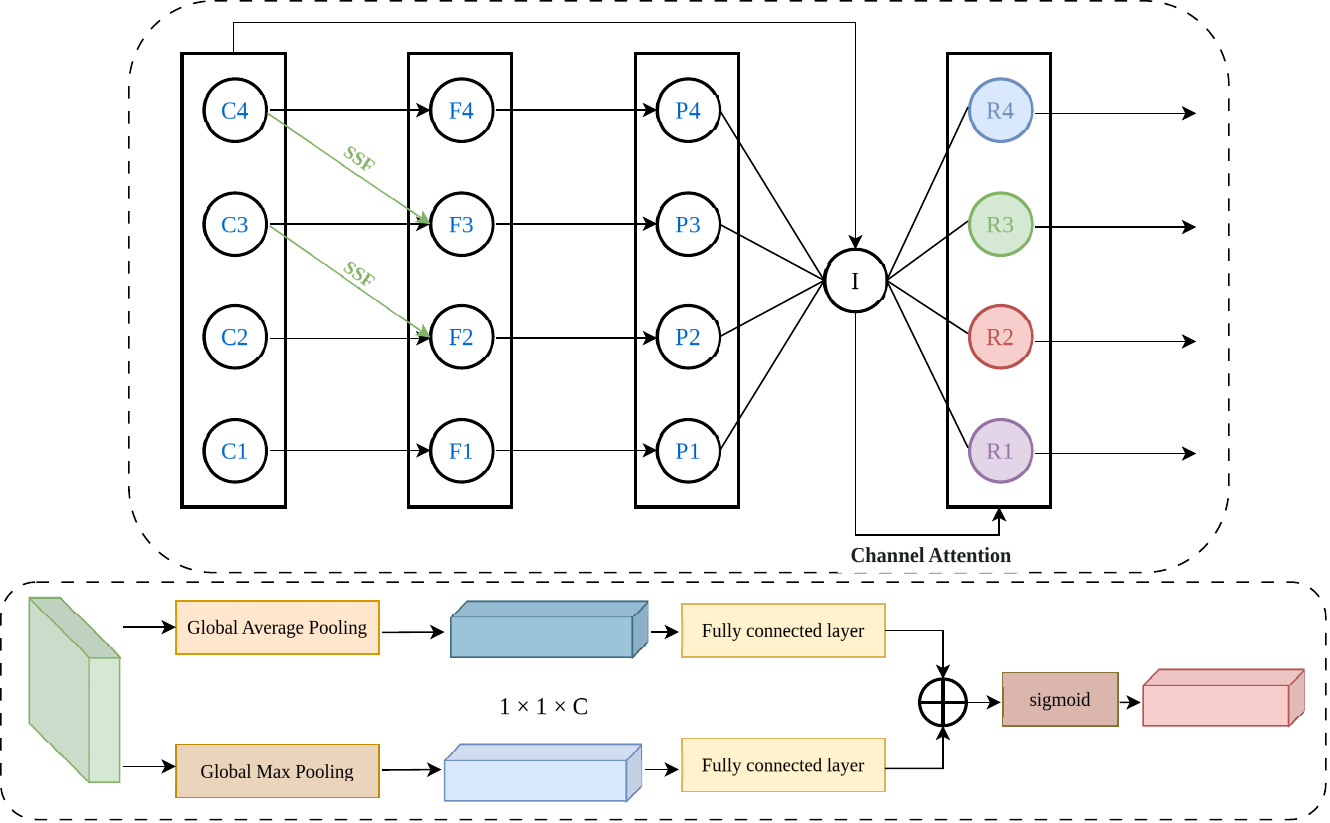}
    \caption{Channel attention-guided FPN (SSF for subpixel skip fusion)}
    \label{fig:fig6}
\end{figure*}

Feature pyramid network (FPN) effectively addresses the issue of object scale variation by combining low-level background information with high-level semantic information, thereby improving detection accuracy. However, FPN has some limitations. For instance, the 1×1 convolution used in lateral connections during the fusion process for channel reduction can lead to the loss of some semantic information. Additionally, the semantic information from different hierarchical features may vary, and directly using interpolation methods for cross-scale fusion can result in aliasing problems. The mixed features generated in this way may confuse localization and recognition tasks. To tackle these issues, CE-FPN \cite{ref26} employs a subpixel skip fusion method (SSF), which fully utilizes the rich channel information from the original cross-scale features. It also introduces a subpixel context enhancement module applied to the top-level features to extract more feature representations. A channel attention module is designed to optimize the integrated features at each layer, enhancing discriminability and alleviating aliasing problems. The network structure of the channel attention-enhanced feature pyramid based on the CE-FPN concept is illustrated in Figure \ref{fig:fig6}.

The core of subpixel skip fusion (SSF) uses sub-pixel convolution to simultaneously alter the number of channels and the size of the feature map. Sub-pixel convolution is an efficient, fast, and parameter-free method for pixel rearrangement and upsampling. In PyTorch, this functionality is implemented by making use of the  \(PixelShuffle\) operation with a specified upscaling factor. This function rearranges the values of the elements in a tensor, changing both the number of channels and the size of the feature map, transforming the dimensions from \( (B, C \times r^{2}, H, W) \) to \( (B, C, H \times r, W \times r) \), where \( r \) is the upsampling factor, indicating the image enlargement ratio. In Figure 6, \( \{ C_{3}, C_{4} \} \) utilizes SSF to perform upsampling without reducing the number of channels and to fuse with low-resolution images. The upsampling factor \( r \) is set to 2, allowing for a twofold spatial dimension expansion for feature fusion \( \{ F_{2}, F_{3} \} \), thereby enhancing the representational capability of the feature pyramid. The process is expressed mathematically as follows:

\begin{eqnarray}
    F_{i} = 
\begin{cases} 
\varphi(C_{i}) + PS(\bar{\varphi}(C_{i+1})) & i = 2, 3 \\ 
\varphi(C_{i}) & i = 1 
\end{cases}
\end{eqnarray}
where, \( \varphi \) represents a 1×1 convolution operation that reduces the channel dimension, the \(PS\) represent the \(PixelShuffle\) function, and \( \bar{\varphi} \) indicates channel transformation using 1×1 convolution, which indexes the feature pyramid. 

The channel attention modul\((CA)\) extracts feature channel weights by integrating the feature map \( I \), which is then multiplied by the different output features. Specifically, the process involves integrating the feature map \( I \) to obtain context information from two different spatial domains through global max pooling and global average pooling. These are then passed into fully connected layers \(fc\). The final channel attention \(CA(I)\) is computed through element-wise summation and a sigmoid function. The mathematical expression for this process is as follows:

\begin{eqnarray}
CA(x) & = & \sigma\left(fc_{1}(AvgPool(x)) + fc_{2}(MaxPool(x))\right) \\
R & = & CA(I) \odot P_{i}
\end{eqnarray}
where, \( CA(x) \) represents the channel attention function, \( \sigma \) denotes the sigmoid activation function, \(P_{i}\) (\(i\in\{1,4\}\)) denote the input features, \(R\) denote the output features, and the symbol \( \odot \) signifies element-wise multiplication.

\section{Experimental results and analysis}
\subsection{Experimental setup}
All experiments were carried out on a system with an Intel i7-10700k CPU, an NVIDIA GeForce RTX 2080Ti GPU, and 32GB of RAM. The implementation was based on PyTorch 1.8.1 and MMDetection v2.15.0. For this study, the initial learning rate was set to 0.015, with the SGD optimizer configured with a momentum of 0.9 and a weight decay of 0.0004. A cosine annealing learning rate schedule was employed to adjust the learning rate over time, with training lasting 110 epochs to reduce the learning rate and improve model stability gradually.

\subsection{Evaluation index}

The detection accuracy is measured using the AP50 metric, where AP stands for average precision. This metric represents the average accuracy across all categories in multi-category predictions. Specifically, AP50 refers to the area under the Precision-Recall curve when the Intersection over Union (IoU) threshold is set to 0.5. It provides a comprehensive evaluation by considering both precision and recall, thereby reflecting the model's ability to identify instances of a specific class correctly. The formulas for calculating precision and recall are as follows:

\begin{eqnarray}
Precision & = & \frac{TP}{TP + FP} \\
Recall & = & \frac{TP}{TP + FN}
\end{eqnarray}
where, \( TP \) (True Positives) is the number of detected boxes with IoU \textgreater 0.5.
\( FP \) (False Positives) is the number of detected boxes with IoU \textless =0.5, or the number of duplicate detections of the same ground truth (GT).
\( FN \) (False Negatives) is the number of ground truth boxes that were not detected.

\subsection{Experimental results}

\begin{figure}[h!]
    \centering
    \includegraphics[width=0.8\linewidth]{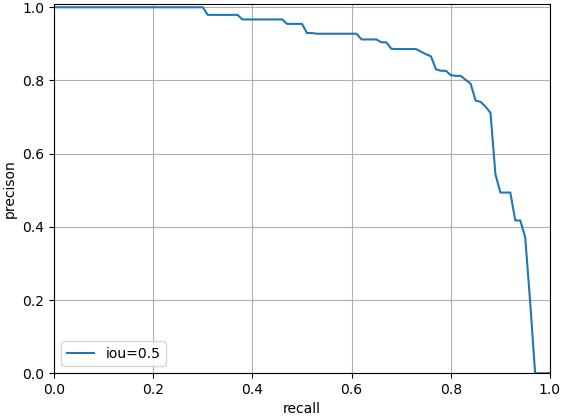}
    \caption{Precision-Recall curve}
    \label{fig:fig7}
\end{figure}

This study trained on the established SRB spectrum dataset. Figure \ref{fig:fig7} illustrates the Precision-Recall (P-R) curve under an IoU threshold of 0.5, indicating that the model demonstrates good detection performance.

\begin{figure*}[h!]
    \centering
    \includegraphics[width=0.9\linewidth]{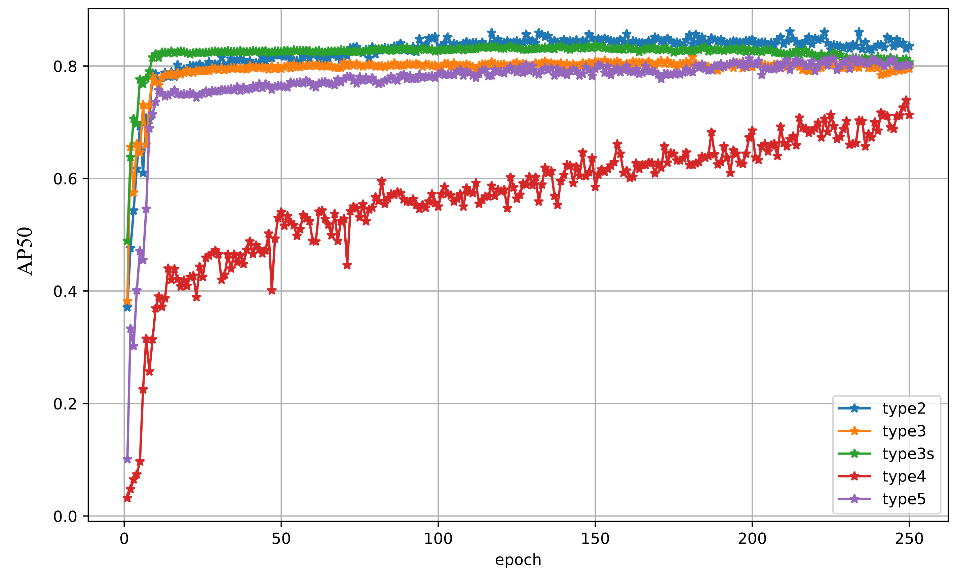}
    \caption{The detection AP50 curves for various types of SRBs events.}
    \label{fig:fig8}
\end{figure*}

Figure \ref{fig:fig8} displays the detection AP50 curves for various types of SRB events. The model exhibited good performance and stability in recognizing these types of events. The accuracy for Type II, III, III-s, and V SRB events was relatively high, allowing for quick convergence, whereas the detection AP50 for Type IV SRBs was comparatively lower.

\subsection{Ablation experiments}
The experiment used TOOD (MobileViTv2 + FPN) as the baseline to verify the effectiveness of each module in the proposed algorithm. Comparative experimental analyses were conducted regarding the scale selection of the output feature maps, the channel attention-enhanced feature pyramid, and other components proposed in this chapter. The metrics for accuracy and recall are presented in Table  \ref{tab:table3}.

\begin{table}[ht]
\caption{Ablation experiments}
\label{tab:table3}
\begin{tabular}{cccccc}
\hline
\hline
MobileViTv2 & FPNLight & CA & Level & mAP@50 & Recall(\%) \\
\hline
\textbf{\checkmark}   &    &    & L5    & 0.754 & 96.0 \\
\textbf{\checkmark}   & \textbf{\checkmark}     &    & L5    & 0.771 & 95.6 \\
\textbf{\checkmark}   & \textbf{\checkmark}   &    & L6    & 0.783 & \textbf{96.1} \\
\textbf{\checkmark}   & \textbf{\checkmark}  & \textbf{\checkmark} & L6  & \textbf{0.799} & 95.1 \\
\hline
\end{tabular}
\end{table}

As seen in Table \ref{tab:table3}, TOOD (MobileViTv2 + FPN) demonstrates a good recall rate; however, its precision is relatively low. After replacing the conventional convolutional networks in the FPN feature pyramid with a lightweight version using depth-separable convolutions, denoted as FPNLight, the mAP@50 increased by 1.7\%. Comparisons were made between feature map scales L5 and L6 fed into the TOOD detection head; selecting six different feature map scales resulted in better model performance, with the mAP@50 value improving by 1.2\% compared to L5. With the addition of channel attention-enhanced FPNLight structure, the mAP@50 value reached 79.9\%. These results provide strong evidence for the effectiveness of the various modules in the proposed algorithm.

\subsection{Comparative experiments with other models}

Table \ref{tab:tab2} presents the experimental comparison results of the proposed model against several other detection models and the TOOD baseline model. These models include YOLOv3, which employs MobileNetv2 as the backbone network, YOLOv5s using K-means anchors (marked as YOLOv5s[a] in Table \ref{tab:tab2}), and YOLOv5s without K-means anchors, utilizing the default sizes (marked as YOLOv5s[b] in Table \ref{tab:tab2}). Additionally, these models also include GFL \cite{ref28}, Swin Transformer, and MobileNetVit SSDLite \cite{ref19}.

\begin{table*}[h!]
\caption{ Comparative experimental results with other models \label{tab:tab2}}
\centering
\resizebox{0.95\linewidth}{!}{ 
\begin{tabular}{lccccccccccc}
\hline
\hline
\multirow{2}{*}{\textbf{Model}} & \multicolumn{6}{c}{\textbf{mAP50@50}} & \multirow{2}{*}{\textbf{Recall(\%)}} & \multirow{2}{*}{\textbf{FLOPs}} & \multirow{2}{*}{\textbf{Params}} \\ 
& \textbf{Avg} & \textbf{\uppercase\expandafter{\romannumeral2}} & \textbf{\uppercase\expandafter{\romannumeral3}} & \textbf{\uppercase\expandafter{\romannumeral3}s} & \textbf{\uppercase\expandafter{\romannumeral4}} & \textbf{\uppercase\expandafter{\romannumeral5}} & & & \\

\hline
YOLOv3         & 0.738 & 0.774 & 0.827 & 0.814 & 0.517 & 0.759 & 89.5  & \textbf{1.69G} & \textbf{3.74M} \\
YOLOV5[a]      & 0.752 & 0.753 & \textbf{0.867} & 0.809 & 0.566 & 0.766 & 69.5  & 7.94G & 7.03M  \\
YOLOV5[b]      & 0.761 & 0.795 & 0.778 & 0.813 & 0.649 & 0.769 & 70.5  & 7.94G & 7.03M \\
GFL            & 0.742 & 0.806 & 0.798 & 0.819 & 0.530 & 0.756 & \textbf{96.6}  & 20.78G  & 32.04M   \\
Swin Transformer & 0.764 & 0.810 & 0.842 & 0.781 & 0.714 & 0.689 & 90.8  & 9.82G  & 28.79M  \\
MobileNetVit-SSDLite & 0.782 & 0.827 & 0.827 & 0.803 & \textbf{0.725} & 0.728 & 92.0  & 2.9G  & 5.32M  \\
ResNet-FPN-TOOD & 0.786 & 0.848 & 0.823 & \textbf{0.822} & 0.672 & 0.764 & 94.9  & 18.08G  & 31.8M   \\
MobileNetVitv2-TOOD(Ours) & \textbf{0.799} & \textbf{0.860} & 0.803 & 0.814 & 0.706 & \textbf{0.810} & 95.1  & 5.14G & 17.6M \\
\hline
\end{tabular}}
\end{table*}

Our proposed model in this study achieved a higher average detection precision and increased recall rates. The baseline model (ResNet-FPN-TOOD) trained from scratch showed good performance in recall rates, but its average precision was relatively low. By utilizing a publicly available ResNet pre-trained model for transfer learning, better experimental results were obtained; however, it performed poorly on Type IV SRB events, with an AP50 of only 67.2\%. Although GFL's recall rate is 0.9\% higher than our proposed model's, our proposed model's average accuracy is 5.7\% higher. In previous studies, MobileNetVit SSDLite \cite{ref19} achieved the best results in multi-category SRB detection, but our proposed model improved the average accuracy by 1.7\% and the recall by 0.2\%.

\subsection{Visualization detection result}
\begin{figure*}[h!]
    \centering
    \includegraphics[width=0.9\hsize]{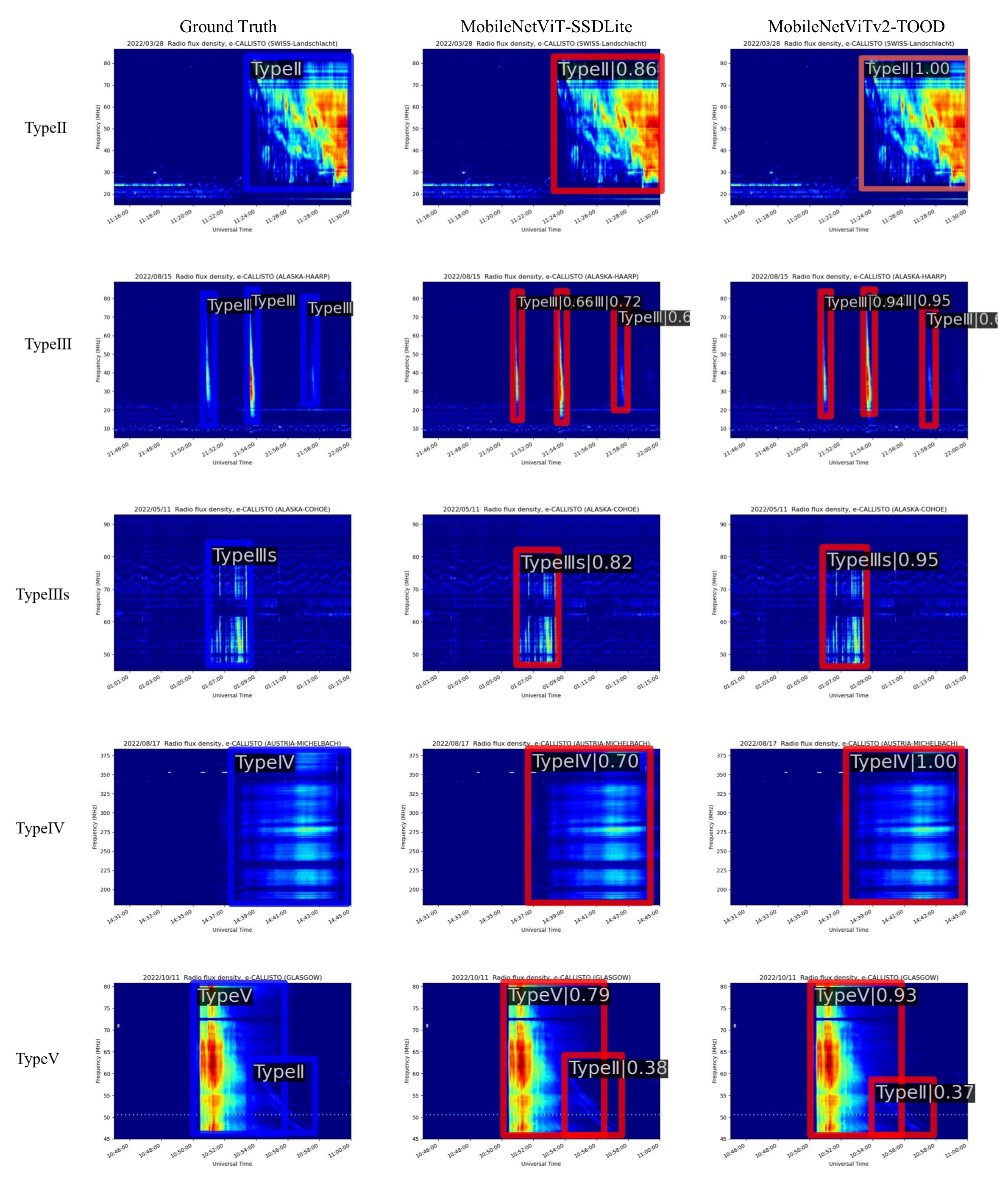}
    \caption{Detection results for Types II, III, III-s, IV, and V SRB events. The abscissa represents time, the ordinate represents frequency, and the intensity of solar radio bursts is reflected by color (the more the color tends to red, the higher the intensity of solar radio bursts).}
    \label{fig:fig9}
\end{figure*}

Figure \ref{fig:fig9}  shows the visualization detection results. The left column of Figure \ref{fig:fig9}  shows the types annotated according to e-CALLISTO. The middle column of Figure \ref{fig:fig9}  shows the results obtained using MobileNetVit SSDLite \cite{ref19}, which is the best result in existing research. The right column of Figure \ref{fig:fig9} shows the results of MobileNetVitv2-TOOD proposed in this paper. In Figure \ref{fig:fig9}, the blue boxes represent the original annotated boxes, and the red boxes provide the localization and classification information for the detected SRBs.

This study improved the detection network by enhancing the feature extraction network with channel attention and refining the task-aligned detection head. Figure \ref{fig:fig9} shows that his study is better regarding detection confidence than the previous best results. 

\begin{figure*}[h!]
    \centering
    \includegraphics[width=0.9\hsize]{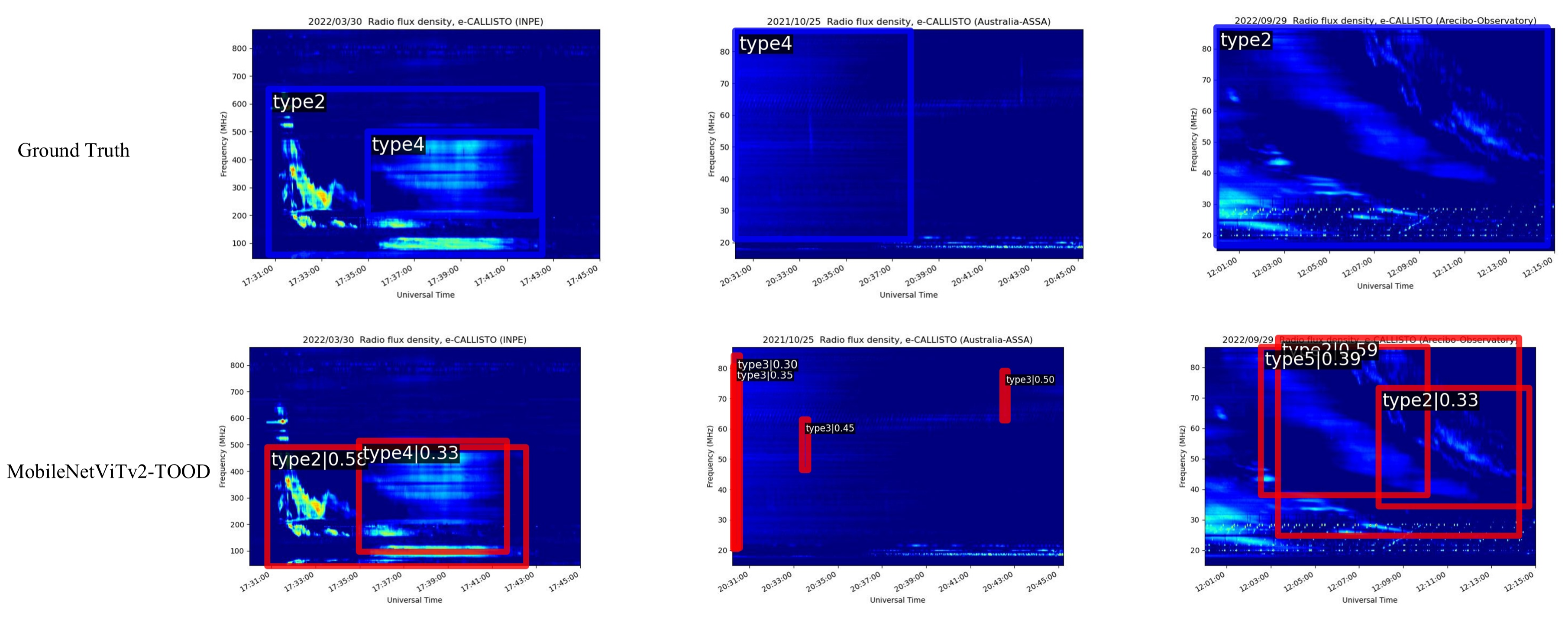}
    \caption{The visualization of the model's limitations}
    \label{fig:fig10}
\end{figure*}

Figure \ref{fig:fig10} shows the performance of our model in complex burst scenarios, which is used to visually demonstrate the limitations of the model. The first row in the figure shows the true bounding boxes, and the second row shows the predicted bounding boxes of the model. It can be seen that in multiple different continuous burst scenarios, especially when the difference between the burst and the background is small and the burst scale is particularly large, the performance of the model is unsatisfactory.

\section{Conclusion}
This study first established a solar radio burst (SRB) spectrum image dataset based on e-CALLISTO, annotating five types of burst events: Type II, Type III, Type IIIs, Type IV, and Type V. Furthermore, Type III was further categorized into single bursts (Type III) and grouped bursts (Type IIIs). Subsequently, an SRB detection model based on the TOOD was proposed. The SRB spectrum images underwent global temporal and spatial feature extraction using the MobileViTv2 backbone network, with features fused by a channel attention-enhanced feature pyramid. Finally, a task-aligned model was employed for classification and localization detection, enabling the automatic recognition and detection of SRB events. The model achieved a detection accuracy of mAP@50 of 79.9\% and a recall rate of 95.1\% on the dataset established in this study. The integration of this model with radio telescopes enables real-time detection of solar radio bursts. Early warnings can be issued during the initial phases of solar radio bursts, thereby effectively reducing the hazards caused by space weather events to various technological systems and human activities.


During the practical application and evaluation of the model, several limitations were identified. A long-tail dataset refers to data exhibiting the characteristic long-tail distribution, where a minority of head classes possess abundant samples while the majority of tail classes suffer from extreme sample scarcity. As demonstrated in Table \ref{table1}, the solar radio burst dataset constructed in this study inherently exhibits a pronounced long-tail distribution characteristic due to the intrinsic patterns of solar eruption occurrences. This imbalanced data distribution leads to universally suboptimal detection performance for tail classes (e.g., Type IV solar radio bursts), a limitation also observed in our proposed model. Furthermore, the model demonstrates compromised detection capabilities in complex solar radio burst scenarios involving concurrent multiple eruption events or intense background noise interference.

To address these limitations, future research will focus on two primary enhancements. For the long-tail data challenge, we plan to implement generative models (e.g., GANs or diffusion models) to synthesize representative samples for tail classes, thereby mitigating the data imbalance and improving tail-class detection robustness. Regarding complex eruption scenarios, we propose to integrate DETR-based architectures (DEtection TRansformer) \cite{ref30} with global attention mechanisms. This adaptation will enable the model to establish comprehensive contextual dependencies across full-spectrum observations, effectively expanding its receptive field and enhancing detection accuracy in multi-event coexistence scenarios. The inherent capability of the transformer for parallelized feature processing is expected to significantly improve computational efficiency in handling spectral interference patterns.

\section*{Acknowledgments}
The authors thank the reviewers for their careful reading and constructive comments. The authors are thankful for the solar radio burst data provided by e-CALLISTO.





\bibliographystyle{unsrt} 
\bibliography{references}

\end{document}